\documentclass[twocolumn,amsmath,amssymb]{revtex4}
\usepackage{graphicx}
\usepackage{color}
\usepackage{booktabs}

\begin{document}
\title{Aging-Induced Dynamics for Statically Indeterminate Systems}
\author{Jr-Jiun Lin$^{1}$, Chi-Chun Cheng$^{1}$, Yu-Chuan Cheng$^{1}$, Jih-Chiang Tsai$^{2,\dagger}$, and Tzay-Ming Hong$^{1,\ast}$}
\affiliation{$^1$Department of Physics, National Tsing Hua University, Hsinchu 30013, Taiwan, Republic of China\\
	$^2$Institute of Physics, Academia Sinica,  Taipei 11529, Taiwan, Republic of China}
\date{\today}
\begin{abstract}

Statically indeterminate systems are experimentally demonstrated to be in fact dynamical at the mesoscopic scale. Take the classic Ladder problem for instance, its weight or frictional force with the wall are found to keep increasing for over $10^5$ s.
We believe that this unanticipated phenomenon is related to the aging effect because the evolution of microcontact area with the wall resulting from the latter follows identical behavior. Finally, simplifications and analytic solutions without invoking detailed material properties are shown to be possible for this statically indeterminate structure when the deformations are small. Since similar findings are also found in  a granular silo and a beam with three support points, we believe this unexpected dynamical variation of normal and frictional forces is intrinsic to all statically indeterminate systems. 
\end{abstract}

\maketitle

Statically indeterminate systems are frequently encountered by architects, engineers, and physicists, such as  a chair/beam with more than three/two legs/supporting points, a ladder leaning against the wall, and the weight of granules in a silo. In these systems, the reaction forces outnumber equilibrium equations and, therefore, are insufficient to be uniquely determined. Physically this implies that they are prone to small perturbations because there are an infinite number of solutions.

The Ladder-Wall problem, being an illustrative example for students learning statics \cite{halliday} as well as a practical case concerning industrial safety \cite{safety, safety2}, has been the target of many discussions in the past. Various textbooks and researchers have considered the problem \cite{principles,statics}, but so far there is no consensus. For instance, Mendelson \cite{mendelson,jon,mech,salu} argued in 1994 that the correct limiting condition is that the  friction force at the top of ladder is equal to the maximum static friction. This was disputed by Gonz$\rm\acute{a}$lez and Gratton \cite{gonz}  two years later, who proposed that the missing condition lies in analyzing flexion.  Dissatisfied with  both assumptions, we decide to verify their validity and come up with a less ad hoc theory.

\begin{figure}[t!]
\centering
        \includegraphics[width=\columnwidth]{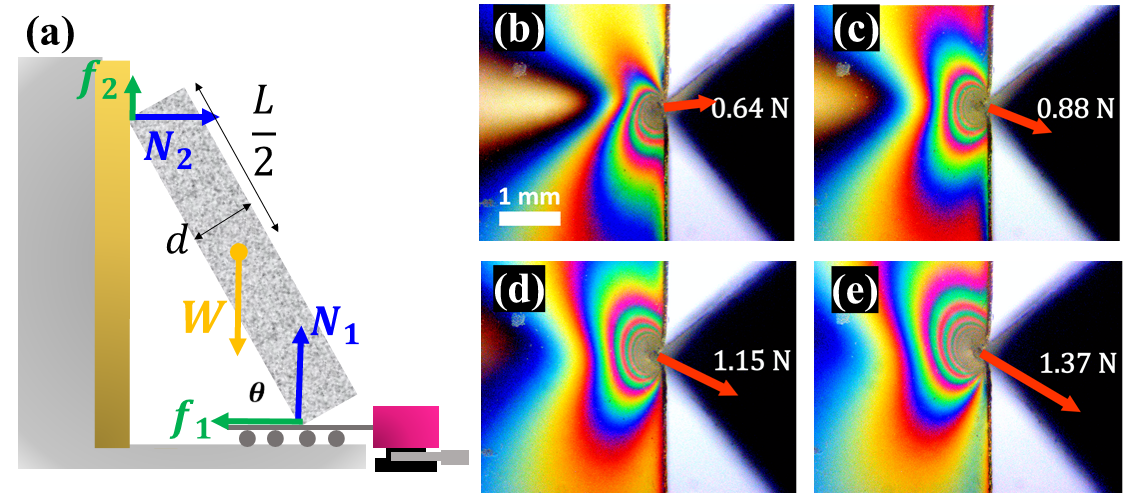}
        
 \caption{(a) Schematics of experimental setup and free body diagram for the ladder/wall system. The PSM-4 attached to the wall (dark yellow stripe) and the force sensor (magenta block) placed behind the floor piece  allows for a multitude of observations. Experiments are conducted with an aluminum bar of $L=10$ cm and $d=4$ cm. (b$\sim$e) Close-up photoelastic images at the contact point. Starting from (b), the micrometer drives forward a distance of $\epsilon\approx 0.2$ mm before the next photo is taken. Force vectors in red arrow are then calculated from the corresponding reading of the force sensor.}
 \label{fig1}
\end{figure}

Consider the beam in  Fig. 1(a). The friction and normal forces are denoted by $f_{1,2}$ and $N_{1,2}$ where subscripts $1$ and $2$ refer to the reaction from the floor and wall. At equilibrium, the equations balancing planar forces and torques can be written as:
\begin{eqnarray}
  \begin{cases}
N_2=f_1\\
N_1+f_2=W\\
\frac{W}{2}(L\cos\theta-d\sin\theta)=f_2L \cos\theta+N_2L \sin\theta
  \end{cases}
 \label{eq1}
\end{eqnarray}
where $W$, $L$, and $d$ denotes the weight, length, and width of the ladder. Clearly these three equations are insufficient to uniquely determine the four unknown reactions. The indeterminacy is resolved by measuring any one of the four unknown forces. As shown schematically in Fig. \ref{fig1}(a), the beam is placed on a movable floor piece where all friction is eliminated by the rollers underneath, and $f_1$ comes solely from the fixed horizontal force sensor\cite{daniel}. The sensor is placed on a micrometer, effectively allowing us to fine-tune the geometry and initial conditions of the setup. 
Aside from the equations and unknowns, the indeterminacy can be displayed by imposing a minute displacement $\epsilon\approx0.2$ mm on the floor piece via adjusting the micrometer so that the change in $\theta$ is  negligible, that results in a substantial variation in the measured forces. The wide range of solutions allowed by the system is further illustrated by the photoelastic material, PSM-4 by Vishay Instruments, attached on the rigid wall\cite{nature,karen}. Figures \ref{fig1}(b)-(e) clearly show that the local stress on PSM-4  redistributes as $f_1$ varies.

One thing that took us by total surprise was that during all experiments with or without the PSM-4, the reading drops monotonically up to $2\times10^4$ seconds. Such a trend reflects a $10\%$ increase in $f_2$, as shown in Fig. 2(a). Is it possible that the ladder may suffer from some mesoscopic slippage? To check it, we estimate that a variation of this magnitude corresponds to roughly $10^{-3}$ rad in change of $\theta$ for $\theta\approx\pi/4$, which should be easily detected by an optical lever. But the result turns out to be negative.

\begin{figure}[ht!]
    \centering
        \includegraphics[width=\columnwidth]{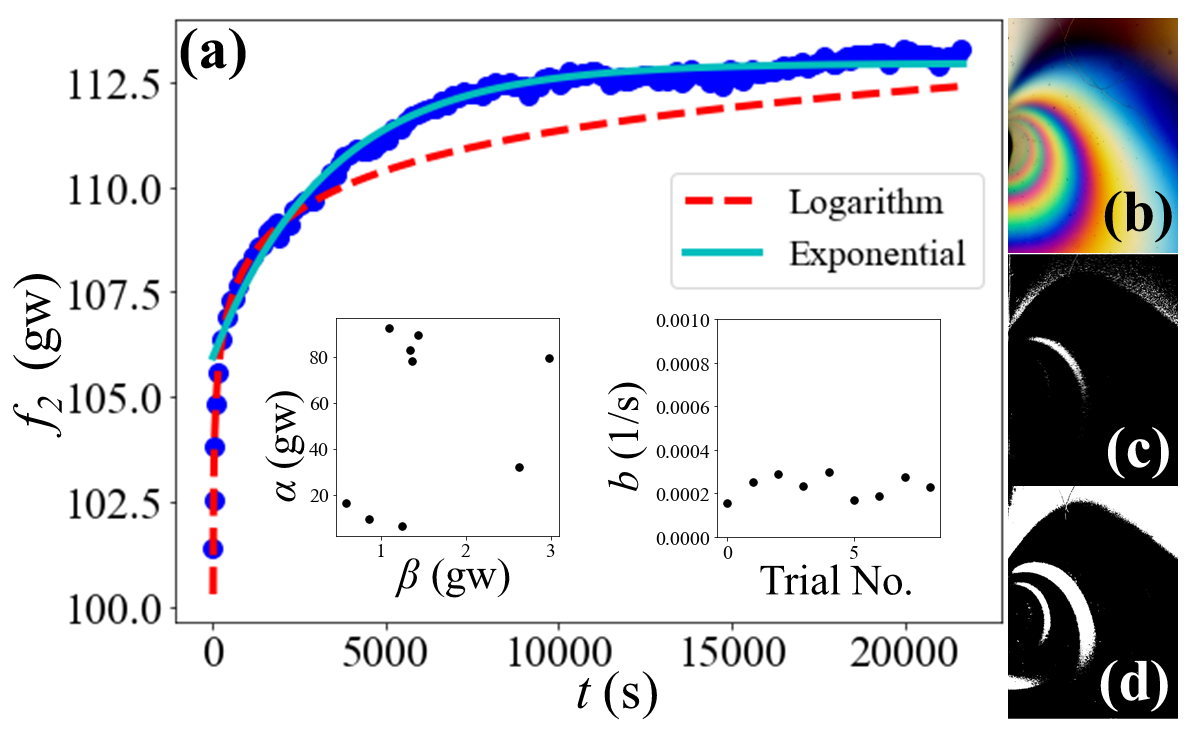}

    \caption{(a) Typical data for $f_2$ at $\theta =\pi /4$. For $t\le 3\times10^3$ s, the data can be fit by the dashed line, $\alpha+\beta\ln t$ with $\alpha=105.1$ and $\beta=1.367$. But if we expand $t$ to $2\times10^4$ s, $a \exp(-bt)+c$ with $a=-7.000$, $b=2.91\times10^{-4}$, and $c=112.953$, becomes a better fit (solid line). The left inset shows that the values of $\alpha, \beta$ fluctuate with different tries and there is no obvious correlation. In contrast, $b\approx (2.32\pm0.52)\times10^{-4}$ in the right inset. (b) Side view of local photoelastic fringes on PSM-4 attached on  the wall at $t=0$. (c, d) Differential image of the same view for contrast at $t=10$ and 500 s.}
\label{fig2}
\end{figure}
For the benefit of later discussions, we convert the $f_1$ data to $f_2$ in Fig. 2(a) that can be nicely fit by $\alpha+\beta\ln t$, characteristic of an aging effect \cite{tabor,f1,f2}. Deviations eventually becomes evident as the data approaches  a flat plateau after roughly $10^4$ s, where $a\exp(-bt)+c$ becomes a better fit.

To elucidate the root source for this unexpected growth in $f_2$, we try to correlate with another measurement. First, a series of photoelastic photos were taken at different times with the setup in Fig. \ref{fig1}(a). The  evolution of interference fringes from Figs. \ref{fig2}(b) to (d) clearly reveals that the distribution of local stresses is as dynamic as $f_1(t)$. To facilitate contrast, the images at subsequent $t>0$ are  turned into grey scale and subtracted from that at $t=0$. The illuminated areas correspond to where fringes have expanded into. Figures 2 (b$\sim$d) show that the expansion is considerable, suggesting a comparable rate of increase for stress in the vicinity of contact points. 

\begin{figure}[htb!]
    \centering
    \includegraphics[width=\columnwidth]{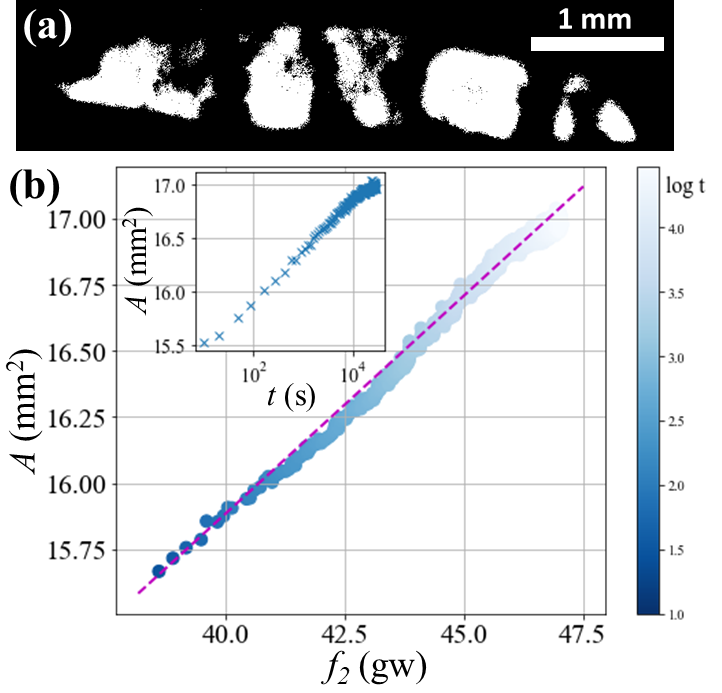}

    \caption{(a) shows binarized images of a rough contact against the glass wall for $\theta =\pi/4$. One pixel roughly corresponds to 10 ${\rm \mu}$m$\times$10 ${\rm \mu}$m. (b) $A$ vs $f_2$ is plotted  over roughly ten hours. Inset shows the logarithmic growth of $A(t)$ in the same trial, consistent with previous studies on aging effect \cite{kilgore,PRR}. }
\label{fig3}
\end{figure}
Meanwhile, a logarithmic growth in static friction coefficient has been observed for various materials including metals\cite{metal1,metal2}, rocks\cite{rock1,rock2} and glassy polymers \cite{poly1,poly2,poly3,poly4,watanabe}. It is established that this growth  stems from the creep of asperities between the two contact surfaces, resulting in an observable increase of real contact area $A$ which is generally only a small fraction of the apparent surface area \cite{tabor,f1,f2}.  Such contacts between asperities are recognized as proxy for frictional force \cite{tabor,kilgore,ruben},  and can be directly observed to the micrometer scale via optical procedures \cite{kilgore,PRR}. We thus deployed a similar experiment by leaning the beam against a smooth glass with a thin rubber padding attached to its end. Light from a LED source is injected from the side of the glass such that it is totally internal reflected except at the contact points where it is scattered and effectively illuminates the area of real contacts to the scale of several microns. The scattered light is captured by the CCD camera on the opposite side of the glass, whose intensity is then binarized for quantitative analysis, as demonstrated by Fig. \ref{fig3}(a).  Results in Fig. \ref{fig3}(b) show that within the time range of $10^4$ seconds, $f_2$ and $A$ maintain a steady linear relation. This confirms the aging of contact area in fact follows an identical time evolution as the increasing $f_2$.

Having presented their  unanticipated variation with time, let us now study how $f_{1,2}$ depend on $\mu_{1,2}$ and $\theta$. In general, the missing condition in statically indeterminate structures, such as the Ladder-Wall problem, requires detailed knowledge of  material properties so that calculations are difficult. In the following we shall propose that simplification  and analytic solutions  for $f_{1,2}(\mu_2,\theta)$ are possible in the case of small deformations. It is heuristic to note that, while the weight affects both bending and stretching, balancing the torques  requires that the perpendicular forces acting on both ends of the beam are equal and fixed. As a result, the extent of bending is uniquely determined by the weight. In contrast, the compression enjoys an additional degree of freedom. In other words, the axial ``squeezing" force on the ladder can take on any arbitrary value as long as Eq. (\ref{eq1}) is satisfied. Therefore, minimizing the compressing force from the wall, $N_2 \cos\theta-f_2\sin\theta$, can fill in as the fourth condition missing from Eq.  (\ref{eq1}), and  the Young's modulus does not  enter the equation. 

However, there are two caveats. First, $N_2 \cos\theta-f_2\sin\theta$ can never be negative based on physical ground. Second, the friction $f_{1,2}$ should never exceed $\mu_{1,2}N_{1,2}$ where $\mu_{1,2}$ denotes the static friction coefficient from floor and wall. In a more enlightening form, the three constraints can each be expressed solely in terms of  $N_1$:
\begin{equation}
N_1\leq\frac{\frac{W}{2}(\cot\theta+\frac{d}{2L})}{\cot\theta-\mu_1}\equiv x_1(\theta,\mu_1)
%\tag{S4.1}
\label{e1}
\end{equation}
\begin{equation}
N_1\geq\frac{W(1+\frac{\mu_2}{2}\cot\theta+\frac{\mu_2d}{2L})}{1+\mu_2\cot\theta}\equiv x_2(\theta,\mu_2)
%\tag{S5.1}
\label{e2}
\end{equation}
\begin{equation}
N_1\geq\frac{W(1+\frac{1}{2}\cot^{2}\theta+\frac{d}{2L}\cot\theta)}{\cot^{2}\theta+1}\equiv x_3(\theta)
%\tag{S6.1}
\label{e3}
\end{equation}
where $x_{1,2}$ are constraints imposed by $\mu_{1,2}$, while $x_3$ comes from $N_2 \cos\theta-f_2\sin\theta\geq 0$.
We are concerned with the boundary beyond which no solution can exist. It is worth noting that only Eq. (\ref{e1}) presents an upper bound. Hence, solutions of $N_1$ cease to exist if and only if $x_1<x_2<x_3$ or $x_1<x_3<x_2$. It is then clear that $x_1=x_2$ and $x_1=x_3$ are the critical conditions that verge on instability, each of which can be rearranged into a neater form:
\begin{equation}
\mu_1=\frac{1-\frac{d}{L}\tan\theta}{\mu_2(1+\frac{d}{L}\tan\theta)+2\tan\theta}
\label{eq2}
\end{equation}
and 
\begin{equation}
\mu_1=\frac{1-\frac{d}{L}\tan\theta}{\cot\theta(1+\frac{d}{L}\tan\theta)+2\tan\theta}
\label{eq3}
\end{equation}
In retrospect, these two phase boundaries are determined by $N_2\mu_2=f_2$ and  $N_2 \cos\theta-f_2\sin\theta=0$, rather than $N_1\mu_1=f_1$.  

\begin{figure}
\centering
        \includegraphics[width=\columnwidth]{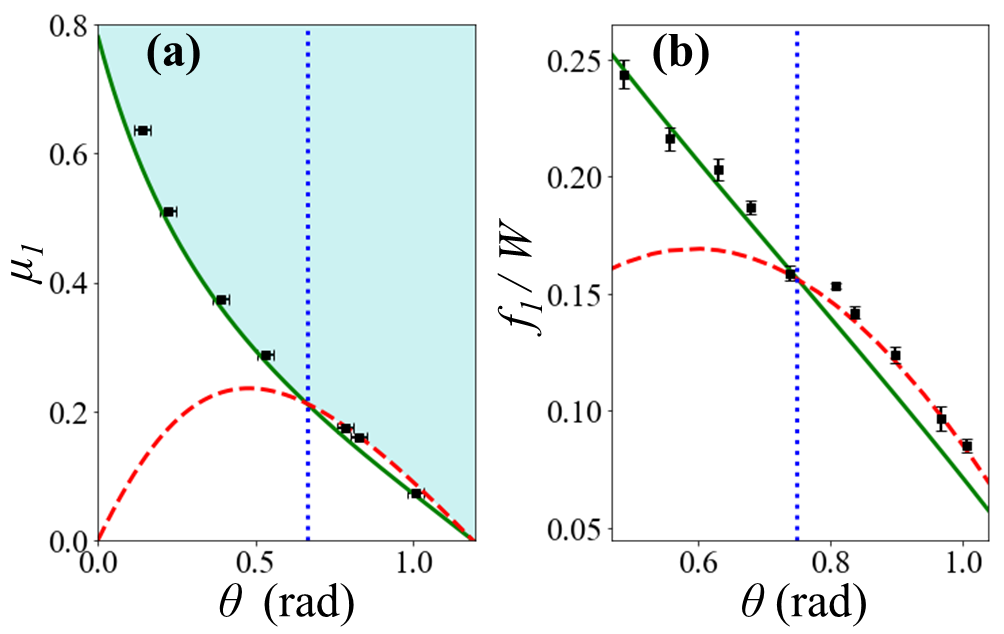}
        
 \caption{(a) Phase diagram for $\mu_1$ vs. $\theta$ with $\mu_2=1.28$. Consistent with the experimental data points, the beam will slip if $\mu_1$ falls below the solid line from Eqs. (\ref{eq2}) when $\theta\le\theta_c=\cot^{-1}(\mu_2)$ marked by the vertical dotted line. Meanwhile, for $\theta\ge\theta_c$ instability is triggered when  $\mu_1$ is below the dashed line from Eq. (\ref{eq3}). (b) Value of $f_1$ is normalized by $W$ and plotted against  $\theta$ with $\mu_1=0.51$ and $\mu_2=1.07$. The solid and dashed lines represent solutions determined by Eqs. (\ref{eq4}) and (\ref{eq5}), respectively.  }
\label{fig4}
\end{figure}

The predictions of Eqs. (\ref{eq2}, \ref{eq3}) are vindicated by the empirical phase diagram  in Fig. \ref{fig4}(a). The beam will slip when its angle falls below either the green solid {\it or} red dashed lines that represent Eq. (\ref{eq2}) and (\ref{eq3}), respectively. Note that Eq. (\ref{eq2}) is equivalent to that derived by previous studies \cite{jon,salu} where $f_2$ is supposed to be at its maximum value. But instead of just a single phase boundary, we argue that the beam is constrained by two distinct requirements across $\theta=\theta_c$. In other words, what is new here is that the phase boundary at $\theta\ge\theta_c$ is governed by Eq. (\ref{eq3}), instead of (\ref{eq2}). A heuristic way to understand this is that  the beam is never in danger of slipping at large $\theta$ and, therefore, has nothing to do with 
$\mu_2 N_2\ge f_2$. As a result, Eq. (\ref{eq3})
or $N_2\cos\theta-f_2\sin\theta\geq0$ steps in and becomes relevant. 

It can be checked that when $\mu_2$ is chosen to be a generic value from 0.4 to 0.7, the red and green lines in Fig. \ref{fig4}(a) become  indistinguishable for $\theta\ge\theta_c$. To discern these two phase boundaries, we purposely attach  a rubber pad to the contact point of the aluminum bar against an aluminum wall which is also padded with another sheet of rubber. Both rubber pads are thin and hard, excluding any  possible deformation of rubber. An apparatus designed as such mimics a real-life ladder, and can yield a rough contact of $\mu_2=1.28\pm0.05$.

Furthermore, in determining the magnitude of $f_{1,2}$ and $N_{1,2}$, the beam prefers to allocate its weight to $f_2$ as much as possible so that $N_2\cos\theta-f_2\sin\theta$ can assume a smaller value to minimize the axial compression energy. Consequently, by requiring the equality in Eqs. (\ref{eq2}) and (\ref{eq3}) to hold separately, the system is no longer indeterminate and explicit solutions  can be acquired as:
\begin{equation}
    f_1=W\Big[-\frac{d}{2L}+\frac{\cot\theta}{2}-\frac{\mu_2(\cot\theta-d/L)}{4(\mu_2+\tan\theta)}\Big]
    \label{eq4}
\end{equation}
for $\theta\le\theta_c$ and
\begin{equation}
    f_1=W\Big[-\frac{d}{2L}+\frac{\cot\theta}{4}-\frac{\sin2\theta}{8}(1-\frac{d}{L})\Big]
    \label{eq5}
\end{equation}
for $\theta\ge\theta_c$. 

In practice, measurements of $f_1$ are largely influenced by initial conditions, such as how the beam is put in contact with the wall, and the logarithmic time dependence that follows. For Fig. \ref{fig4}(b) that checks the correctness of Eqs. (\ref{eq4}, \ref{eq5}), our protocol is to collect data at roughly $10^3$ s. With an automated electromagnet, the beam is kept several millimeters away from the wall before each trial. By shutting down the electromagnet, we can control when to drop the beam gently onto its leaning position at the designated angle. 

\begin{figure}
\centering
\includegraphics[width=\columnwidth]{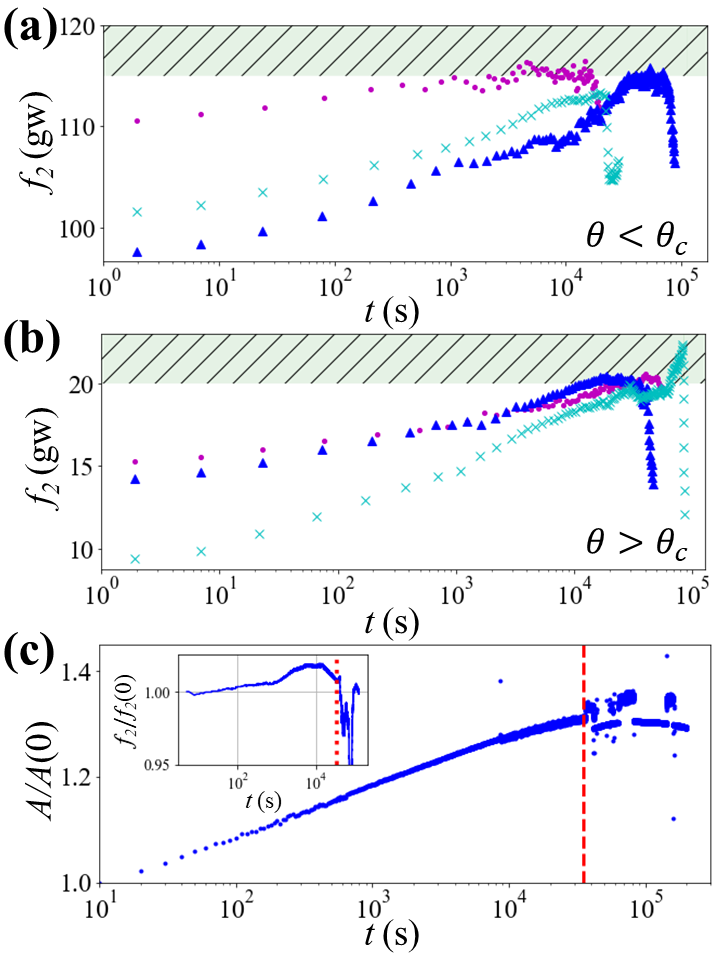}
        
 \caption{(a) Evolution of $f_2$ for three distinct trials with identical setup and conditions is plotted against $\log t$ with $\theta$=0.54 rad. Shaded bar indicated the maximum value of $f_2$ beyond which $f_2\le N_2\mu_2$ will be violated. However, note that $\mu_2$ may increase slightly with $t$ \cite{f1}. Data points far beyond the steep drop is not shown for clarity. (b) Three trials for $\theta$=0.96 rad. Shaded bar indicates the forbidden regime where $N_2\cos\theta-f_2\sin\theta\le 0$. (c) As the measured $f_2$ drops spontaneously in the inset, $A$ underwent noticeable fluctuation. The dashed vertical line marks and shows that the onset of these two phenomena coincide.}
 \label{fig5}
\end{figure}

Can we combine the data in Fig. \ref{fig2} with Eq. (\ref{eq4}) and claim to have obtained and confirmed\cite{metal1,metal2,rock1,rock2,poly1,poly2,poly3,poly4,watanabe} the logarithmic growth for $\mu_2(t)$? The answer is negative. A quick reason is that this argument apparently does not work for the large-angle solution in Eq. (\ref{eq5}) since it consists only of constants. As a result, Eqs. (\ref{eq4}, \ref{eq5}) should be treated as stationary solutions after the indeterminate system already reaches equilibrium.

Still, there is the issue of a time-dependent $\mu_2$ since the normal load on the contact surface is constantly changing. Past experiments with constant load generally find the slope  $d\mu_2/d\log(t)$ in the order of $10^{-2}$ for various materials \cite{f1}, according to  which the increment of the maximum $f_{2}$ in Fig. \ref{fig5}(a)  should be no more than $5\%$ after several hours. But, since the normal force, $N_2$, decreases with time, an aging $\mu_2$ is expected to play a lesser role in explaining the time evolution of $f_{1,2}$ than the indeterminate nature of this system. The match between experimental results and theory in Fig. \ref{fig4}(b) offers convincing evidence that minimizing the compression energy indeed plays a decisive role for the statically indeterminate Ladder-Wall system.

%Static friction usually emerges only as a reaction upon external shear, and the effect of aging rough surface manifests in the increase of maximum shear before slipping. In contrast, a statically indeterminate system allows a range of possible solutions for the strength of friction, and presents a physical value according to the tendency governed by the system's energy. In the Ladder-Wall case, the logarithmic aging surface at the wall takes on a proportional amount of shear as time passes, minimizing the energy of the beam, and thus an increase in $f_2$ (reflected by the declining $f_1$) is observed.

The fact that the $\ln t$ fitting fails at $t\approx 10^4$ s in Fig. \ref{fig2}(a), much sooner than other aging systems, e.g., $10^6$ s for a crumpled sheet \cite{nagel}, is due to the upper cut-off in Fig. \ref{fig5}(a, b) set  respectively by $f_2=N_2\mu_2$ and $N_2\cos\theta-f_2\sin\theta=0$ at small and large $\theta$. Although the initial value of $f_2$ varies with different trials, they all observe the similar logarithmic growth at short time and level off before their maximum value is reached. 

%In the experiment illustrated in Fig. \ref{fig5}(a), repeated trials at $\theta\leq\theta_c$ are carried out, where measurements last up to one day. The Y-intercept and slope indeed alter from trial to trial, despite the lack of noticeable change in experimental conditions. Logarithmic growth comes to a halt near the point of maximum friction, where the time evolution starts to deviate from the initially steady trend. Fig. \ref{fig5}(b) depicts trials done with $\theta\geq\theta_c$, which exhibits similar features. 
 
 We notice a common feature  in Fig. \ref{fig5} that we do not yet understand. Namely, upon reaching the upper limit $f_2$ will fluctuate for several hours before a precipitous drop. 
 %Intuitively, clashing with the maximum friction seem to indicate spontaneous slipping at the wall, which is however forbidden by friction at the floor. Such anomaly is thus conjectured to be tied with mesoscopic episodes taking place at the contact point.
 Simultaneously, the contact area begins to fluctuate vehemently to the point it appears discontinuous at times, as shown in Fig. \ref{fig5}(c). 
 
To test the generality of our conclusions, we also studied a horizontal beam with three supports and found the measured weight on each support displays an identical logarithmic relaxation up to $10^4$ s \cite{sm}. Another system that we checked is a popular topic in soft matters and normally does not remind people of being statically indeterminate - a granular silo. By measuring the pressure exerted by the granules on the  bottom plate that is detached from the silo \cite{reverse}, we again found the logarithmic relaxation for $10^4$ s \cite{sm}. Had we included the weight of the silo, the total weight certainly would have been a constant. But,  the granules-silo becomes analogous to the ladder-wall arrangement when the bottom plate is separated from the silo.

In conclusion,  we reported a surprising time-dependent behavior for the friction and normal forces in the Ladder-Wall problem - a textbook example for statically indeterminate systems. This phenomenon is shown to be correlated to local dynamics at the contact between the ladder and wall. Specifically, both the forces and contact area obey the same logarithmic relaxation, typical for the aging  effect. Against the general impression that the solution to a statically indeterminate system requires detailed material properties and are very difficult to calculate,
we claimed that simplifications and analytic solutions are in fact possible  as long as the deformations are small. Vindicated by experiments, the predictions and derivations of our approach are heuristic and only make use of fundamental physical constraints. 

We gratefully acknowledge technical assistance from Ke-Yang Hsiao and Li-Min Wang and financial support from MoST in Taiwan under Grants No. 105-2112-M007-008-MY3 and No. 108-2112-M007-011-MY3.

\end{document}